\documentclass[10pt,prd,twocolumn,nofootinbib]{revtex4}

\def\mysection#1{{\bf #1.} }

\usepackage{amsmath,amssymb,verbatim,mathrsfs,array,layout,textcomp,amssymb,latexsym, slashed,graphicx,bbm}

\usepackage{graphicx}
\usepackage{amsmath}
\usepackage{amssymb}
\usepackage{appendix}
\usepackage{verbatim,mathrsfs,array,layout,textcomp,latexsym, slashed}

\usepackage{amsmath, amssymb, rotating}
\usepackage{hyperref}
\usepackage[usenames,dvipsnames]{color}

\usepackage{bm}

\newcommand{\beq}{\begin{eqnarray}}
\newcommand{\eeq}{\end{eqnarray}}

\def\beqa{\begin{eqnarray}}
\def\eeqa{\end{eqnarray}}
\newcommand{\no}{\nonumber}
\newcommand{\bv}{\left(\begin{array}{c}}
\newcommand{\ev}{\end{array}\right)}
\newcommand{\bmtwo}{\left(illustrated in\begin{array}{cc}}
\newcommand{\bmthree}{\left(\begin{array}{ccc}}
\newcommand{\emn}{\end{array}\right)}
\newcommand{\bmtwoc}{\left\{\begin{array}{cc}}
\newcommand{\bmthreec}{\left\{\begin{array}{ccc}}
\newcommand{\emnc}{\end{array}\right\}}
\newcommand{\ba}{\begin{array}}
\newcommand{\ea}{\end{array}}

\newcommand{\be}{\begin{equation}}
\newcommand{\ee}{\end{equation}}

\newcommand{\s}{{\rm s}}
\newcommand{\p}{{\rm p}}
\newcommand{\px}{{\rm p}_x}
\newcommand{\py}{{\rm p}_y}
\newcommand{\pz}{{\rm p}_z}

\newcommand{\vecr}{\mathbf{r}}

\newcommand{\vecq}{\mathbf{q}}
\newcommand{\veck}{\mathbf{k}}

\newcommand{\vecl}{{\bm \ell}}
\newcommand{\vecv}{\mathbf{v}}

\newcommand{\eV}{\text{ eV}}

\def\lsim{\mathrel{\rlap{\lower4pt\hbox{\hskip1pt$\sim$}}
     \raise1pt\hbox{$<$}}}         
\def\gsim{\mathrel{\rlap{\lower4pt\hbox{\hskip1pt$\sim$}}
     \raise1pt\hbox{$>$}}}         

\begin{document}

\font\mini=cmr10 at 0.8pt

\title{
Directional Detection of Dark Matter with Two-Dimensional Targets
}

\author{Yonit Hochberg${}^{1,2,3,4}$}\email{yonit.hochberg@cornell.edu}
\author{Yonatan Kahn${}^5$}\email{ykahn@princeton.edu}
\author{Mariangela Lisanti${}^{5}$}\email{mlisanti@princeton.edu}
\author{Christopher G. Tully$^{5}$}\email{cgtully@princeton.edu}
\author{Kathryn M. Zurek${}^{1,2}$}\email{kzurek@berkeley.edu}
\affiliation{${}^1$Ernest Orlando Lawrence Berkeley National Laboratory, University of California, Berkeley, CA 94720, U.S.A.}
\affiliation{${}^2$Department of Physics, University of California, Berkeley, CA 94720, U.S.A.}
\affiliation{${}^3$Department of Physics, LEPP, Cornell University, Ithaca, NY 14853, U.S.A.}
\affiliation{${}^4$Racah Institute of Physics, Hebrew University of Jerusalem, Jerusalem 91904, Israel}
\affiliation{${}^5$Department of Physics, Princeton University, Princeton, NJ 08544, U.S.A.}

\begin{abstract}
   We propose two-dimensional materials as targets for direct detection of dark matter. Using graphene as an example, we focus on the case where dark matter scattering deposits sufficient energy on a valence-band electron to eject it from the target.  We show that the sensitivity of graphene to dark matter of MeV to GeV mass can be comparable, for similar exposure and background levels, to that of semiconductor targets such as silicon and germanium.  Moreover, a two-dimensional target is an excellent directional detector, as the ejected electron retains information about the angular dependence of the incident dark matter particle.
    This proposal can be implemented by the PTOLEMY experiment, presenting for the first time an opportunity for directional detection of sub-GeV dark matter.
\end{abstract}

\maketitle

\section{Introduction}\label{sec:intro}
The Weakly Interacting Massive Particle (WIMP) is currently the dominant theoretical paradigm for dark matter (DM), and has guided experimental search efforts in recent decades.  Direct detection experiments, which search for DM-nucleus collisions, are currently targeting the WIMP parameter space~\cite{Aprile:2012nq,Agnese:2013jaa,Agnese:2014aze,Akerib:2015rjg,Angloher:2015ewa,Agnese:2015nto,Aprile:2016wwo}.  However, null results from these searches motivate renewed consideration for a broader range of DM models.  One possibility involves DM particles below the $\sim$GeV scale, which arise in a variety of theory scenarios~\cite{Mohapatra:2000qx,Mohapatra:2001sx,Boehm:2003hm,Strassler:2006im,Pospelov:2007mp,Hooper:2008im,Feng:2008ya,
Kaplan:2009ag,Zurek:2013wia,Hochberg:2014dra,Hochberg:2014kqa,DAgnolo:2015koa,Kuflik:2015isi,Pappadopulo:2016pkp}.  Current direct detection experiments lose sensitivity to sub-GeV DM because the nuclear recoil energy is too small to be detected.  However, DM with mass below a target nucleus deposits a greater fraction of its kinetic energy on an electron than a nucleus, making electrons a favorable target for light DM detection.

Consider the case of MeV-scale DM, which carries about an eV of kinetic energy, enough to excite atomic electrons after scattering~\cite{Essig:2011nj}. The first limits on such processes have been set using data from the Xenon10 experiment~\cite{Essig:2012yx}, with recent work extending this analysis to Xenon100~\cite{Essig:2017kqs}.  The energy gap for electronic excitations in noble gases is $\sim$10~eV, which places a lower bound on the DM mass that can be probed with these methods. However, a smaller energy deposit can up-scatter valence electrons in semiconductors with band gaps $\sim$1~eV~\cite{Essig:2011nj,Graham:2012su}.  As a result, semiconductor targets are more sensitive to DM in the 1--10~MeV mass range~\cite{Lee:2015qva,Essig:2015cda}.  Superconducting targets with $\sim$meV energy gaps are capable of reaching $\sim$keV masses~\cite{Hochberg:2015pha,Hochberg:2015fth}.

This Letter proposes an alternative approach using two-dimensional (2D) materials as targets.  In this setup, an incident DM particle can deposit sufficient energy on a valence electron to eject it from the target.  The energy and direction of the recoiling electron is then directly measured with a combination of position measurements, time-of-flight, and energy deposition in a calorimeter.  This is in contrast to scattering in bulk targets, where the scattered particle (nucleus or electron) produces secondary excitations before measurement \cite{Agnese:2013jaa,Agnese:2015ywx,Lee:2015qva,Essig:2015cda}, erasing the initial directional information in the scattering.
Using 2D targets, DM masses down to the MeV scale can be probed if the energy required to eject the electron is a few eV.

Most importantly, 2D targets allow one to measure the direction of the incoming DM because the differential cross section for the outgoing electron is peaked in the forward direction.  The lattice structure of the target can even yield diffraction patterns in the electron angular distribution for certain kinematics.  Directional detection has long been recognized as a powerful tool in the study of DM, both as a discriminator against background sources and also because it leads to a daily modulation of the signal rate~\cite{Spergel:1987kx}.  There are currently no feasible proposals for directional detection of sub-GeV DM~\cite{Mayet:2016zxu}, making the use of 2D targets a powerful tool in pushing sensitivities to lower DM masses. We will describe a potential experimental realization using the PTOLEMY experiment~\cite{Betts:2013uya}.

\section{Dark Matter Scattering in Graphene}\label{sec:DMscat}  As a concrete example of a 2D target, we focus on monolayer graphene, which is especially convenient because analytic solutions for the electron wavefunctions in the tight-binding approximation are tractable due to the symmetries of the lattice~\cite{dresselhaus}. We compute the DM scattering rate here, and in Section~\ref{sec:dir} we show that the direction of the scattered electron retains a strong directional correlation with the DM direction.

Monolayer graphene consists of carbon atoms arranged in a two-dimensional honeycomb lattice. The distance between neighboring carbon atoms is $a = 0.142 \ {\rm nm}$. The lattice is built from two distinct triangular sub-lattices.  Four out of the six electrons of a carbon atom are valence electrons, occupying $(2\s)(2\p)^3$ orbitals.\footnote{The core $1\s$ electrons have binding energies of several hundred~eV and contribute negligibly to the scattering rate.} The $2\s$ orbital becomes `hybridized' with the in-plane $\px$ and $\py$ orbitals, such that the energy eigenstates (called $\sigma$ bonds) are linear combinations of $2\s$, $2\px$, and $2\py$. The out-of-plane $\pz$ orbitals remain unhybridized and form covalent bonds, called $\pi$.  We outline the important features of the unhybridized $\pi$ electron wavefunction here, relegating further details and a discussion of the $\sigma$ electrons to Appendix \ref{app:Wavefunctions}.

Within the tight-binding model, we approximate the wavefunction by a sum over nearest neighbors, corresponding to four lattice sites.  The Bloch function for a $\pi$ electron is given by
\be
\Psi_\pi(\vecl,\vecr) \approx {\cal N}_\vecl \, \left(\phi_{2\pz}(\vecr) +   e^{i \varphi_\vecl} \, \sum_{j=1}^3 e^{i \vecl \cdot \textbf{R}_j}\, \phi_{2\pz}(\vecr - \textbf{R}_j)  \right)
\label{eq:wavefunction}
\ee
for lattice momentum $\vecl = (\ell_x, \ell_y) \in {\rm BZ}$ in the Brillouin zone.  Here, ${\cal N}_\vecl$ is a normalization constant, $\textbf{R}_j$ are the nearest-neighbor vectors, and $\varphi_\vecl$ is an $\vecl$-dependent phase.  We take a hydrogenic orbital for the $2\pz$ wavefunction of carbon,
\be\label{eq:orb}
\phi_{2\pz}(\vecr) = {\cal N} \,a_0^{-3/2}\, \frac{r}{a_0}\, e^{- Z_{\rm eff} r/2 a_0}\, \cos \theta\,,
\ee
where $a_0$ is the Bohr radius and ${\cal N}$ is the normalization.  The effective nuclear charge $Z_{\rm eff} \simeq 4.03$ is chosen to fit the numerical solution for the overlap between adjacent $2\pz$ orbitals.  The Fourier transform of Eq.~\eqref{eq:wavefunction} is
\be
\widetilde \Psi_\pi(\vecl, \veck) = {\cal N}_\vecl \left(1 + e^{i \varphi_\vecl} \, f\left(\vecl + \veck\right) \right) \widetilde \phi_{2\pz}(\veck),
\ee
where $\veck$ is the momentum conjugate to $\vecr$,  $f(\vecl + \veck) = \sum_{j=1}^3 e^{i (\vecl + \veck) \cdot \textbf{R}_j}$ is a sum of phase factors, and the Fourier transform of the atomic orbital is well-approximated by
\be
\label{eq:Phi2pz}
\widetilde \phi_{2\pz}(\veck) \approx \widetilde{{\cal N}} \, a_0^{3/2} \frac{a_0 \,k_z }{\left( a_0^2 \, |\veck|^2+ \left(Z_\text{eff}/2\right)^2\right)^3}
\ee
with normalization $\widetilde{{\cal N}}$.

Analytic forms for the $\sigma$ electron wavefunctions are also possible to derive, but are more complicated than their $\pi$ counterparts because the coefficients of the basis orbitals must be computed by diagonalizing a $6 \times 6$ Hamiltonian. The $\pi$ ($\sigma_1$) electrons have binding energies $\sim$0--6~(13--18)~eV.

\begin{figure*}[t]
\begin{center}
	\includegraphics[width=\textwidth]{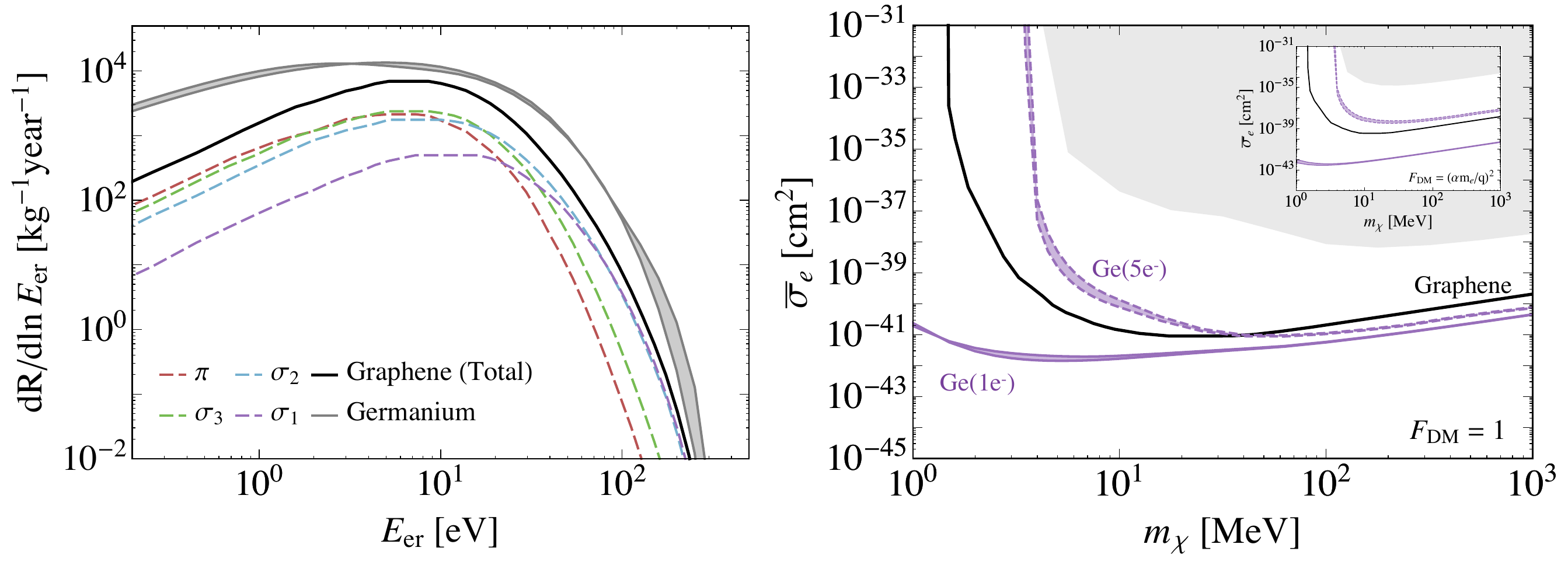}
\end{center}
\vspace{-0.3cm}
 \caption{(\emph{left}) Differential rate for a 100~MeV DM particle scattering off an electron in graphene with $\bar{\sigma}_e = 10^{-37}$~cm$^2$ and $F_\text{DM}(q) = 1$.  The solid black line denotes the total rate, while the dashed lines show the contributions for electrons in the individual $\pi$ and $\sigma$ bands.  For comparison, the differential rate for germanium, taken from Ref.~\cite{Lee:2015qva}, is shown in gray; the band denotes the variation due to scattering off the 4s or 4p valence electron.  (\emph{right})  Expected background-free 95\% C.L. sensitivity for a graphene target with a 1-kg-year exposure (black). Also plotted are the analogous curves for germanium~\cite{Lee:2015qva} with 1-electron (solid purple) and 5-electron (dashed purple) thresholds including the variation due to 4s/4p bands, and exclusions from Xenon10 and Xenon100 \cite{Essig:2017kqs} (shaded gray). We consider both heavy-mediator exchange, which leads to $F_{\rm DM}(q)=1$, and light-mediator exchange, $F_{\rm DM}(q)= (\alpha m_e/q)^2$ (inset).
  \label{fig:diffR}}
\vspace{-0.4cm}
\end{figure*}

If the scattered electron is ejected from the material after scattering, then its final-state wavefunction is well-modeled by a plane wave \cite{arpes}.  The initial-state wavefunction corresponds to an electron in any of graphene's four valence bands. The cross section for a DM particle of mass $m_\chi$ and initial velocity $\vecv$ to scatter off an electron in band $i= \pi,\sigma_1,\sigma_2,\sigma_3$ with lattice momentum $\vecl$ is then
\beq
 \label{eq:vdsigma}
v \, \sigma_i(\vecl)  \!\!&=& \!\!\frac{\bar{\sigma}_e}{\mu_{e\chi}^2}\int \frac{d^3 k_f}{(2\pi)^3}\,\frac{d^3 q}{4 \pi} \,\left|  F_\text{DM}(q)\right|^2 \,\left|\widetilde{\Psi}_i(\vecl, \vecq - \veck_f)\right|^2  \no \\
\!&\times& \!\!\!  \,\delta\!\left(\frac{k_f^2}{2m_e} + E_i(\vecl) + \Phi  + \frac{q^2}{2m_\chi} \!- \mathbf{q}\cdot \mathbf{v}\right) \, ,
 \eeq
where $-E_i(\vecl)$ is the band energy, $m_e$ is the electron mass, $\veck_f$ is the final electron momentum, $\vecq$ is the momentum transfer (\emph{i.e.},\ the outgoing DM has momentum $m_\chi \mathbf{v} - \vecq$), and $\mu_{e \chi}$ is the DM-electron reduced mass. Note that $\vecq$ and $\veck_f$ are independent scattering variables since the initial bound-state wavefunction is an energy eigenstate but not a momentum eigenstate, with Fourier components at all $\veck$ values. For the same reason, Eq.~(\ref{eq:vdsigma}) contains only a single delta function enforcing energy conservation, with no corresponding delta function for momentum conservation. $\Phi \simeq 4.3 \eV$ is the work function of graphene~\cite{arpes}, defined as the energy difference between the Fermi surface and the vacuum.\footnote{The work function is not an intrinsic property of graphene, and can be manipulated with a suitable choice of substrate; see \emph{e.g.}, Ref.~\cite{yuan2015engineering}.}  Following Ref.~\cite{Essig:2011nj}, we define
\be
\label{eq:sigdef}
\bar\sigma_e\equiv \frac{\mu_{e \chi}^2}{16\pi m_\chi^2 m_e^2}\overline{\left|{\cal M}_{e \chi}(q)\right|^2} \Big|_{q^2=\alpha^2 m_e^2}\, ,
\ee
with $\overline{\left|{\cal M}_{e \chi}(q)\right|^2}$ the spin-averaged amplitude, to be the scattering cross section for DM off a free electron with $q=\alpha\, m_e$.  The momentum dependence of the matrix element is then absorbed into the DM form factor
%
$\label{eq:Fdm}
F_\text{DM}(q) =\overline{\left|{\cal M}_{e\chi}(q)\right| }/\overline{\left|{\cal M}_{e\chi}(\alpha\, m_e)\right|}\,.
$
We do not include the so-called Fermi factor, which enhances the rate at low recoil energies due to the distortion of the outgoing electron wavefunction by the Coulomb field of the nucleus. This factor is significant for bulk materials, but negligible for a 2D material for two reasons: the ionized electron energy must be high enough to overcome the work function, and the ionized electron travels single-atom distances and thus spends little time in the vicinity of the nucleus.

To obtain the total rate per unit time and detector mass, we must integrate Eq.~\eqref{eq:vdsigma} over all $\vecl \in {\rm BZ}$ and all incoming DM velocities, then sum the contributions from the four valence bands:
\be
 \label{eq:fullR}
 R = 2 \,\sum_{i= \pi, \sigma_{1,2,3}} \frac{\rho_\chi}{m_\chi} \, N_\text{C} \, A_\text{uc} \int  \frac{d^2 \ell}{(2\pi)^2} \, d^3 v \, g(\mathbf{v})\;  v \,\sigma_i(\vecl) \, ,
 \ee
where $g(\mathbf{v})$ is the lab-frame DM velocity distribution, $A_\text{uc} = 3\sqrt{3}a^2/2$ is the area of the unit cell, $N_\text{C} \simeq5\times 10^{25}\;{\rm kg}^{-1}$ is the density of carbon atoms in graphene, and $\rho_\chi \simeq 0.4\;{\rm GeV}/{\rm cm}^3$ is the local DM density \cite{Read:2014qva}.  The factor of two in Eq.~\eqref{eq:fullR} accounts for the degenerate spin states in each band.

The kinematics of the scattering process dictate that there is a minimal DM velocity required to eject an electron of momentum $k_f$ from the target via a momentum transfer $q$:
\be\label{eq:vmin}
v_{\rm min}^i(\vecl, k_f, q) = \frac{E_\text{er} + E_i(\vecl) + \Phi}{q} + \frac{q}{2m_\chi}\,,
\ee
where $E_\text{er}\equiv k_f^2/2m_e$. We assume the Standard Halo Model (SHM) \cite{Drukier:1986tm} for $g(\mathbf{v})$, with $v_0 \simeq 220\;{\rm km}/{\rm s}$ \cite{McMillan:2009yr} and $v_{\rm esc} \simeq 550\;{\rm km/s}$ \cite{Smith:2006ym} (relative to the Galactic frame). For an electron at the Fermi surface with $E_i(\vecl) = 0$, the minimum $q$ needed for $v_{\rm min} = 782 \ {\rm km}/{\rm s}$, the largest possible lab-frame DM velocity in the SHM, is $q_{\rm min} \simeq 1.6 \ {\rm keV}$. Comparing this with the inverse atomic spacing $2\pi/a \simeq 8.7 \ {\rm keV}$, we see that all kinematically allowed scattering is localized to only a few unit cells, with most confined to a single one. We have verified numerically for the $\pi$ band that the nearest-neighbor approximation made in Eq.~\eqref{eq:wavefunction} is sufficient.

Approximating the SHM velocity distribution as isotropic, $g(\mathbf{v}) = g(v)$, Fig.~\ref{fig:diffR} (left) shows the differential scattering rate for a 100~MeV DM particle. The total rate (solid black line) is comparable to that for a germanium target (gray band). The contributions from the individual $\pi$ and $\sigma$ electrons are indicated by the dashed lines.  Although electrons in the lowest two $\sigma$ bands contribute the least at low recoil energies,
they dominate at higher recoil energies.  This is because the $\sigma_{1,2}$ bands are mostly $2\s$ and therefore have a larger spread in momentum.

In the right panel of Fig.~\ref{fig:diffR}, we show the 95\% one-sided Poisson C.L.\ expected reach (3.0 events) after 1-kg-year exposure of a graphene target, assuming a zero-background experiment. The reach is plotted for form factors of both heavy and light mediators, $F_{\rm DM} (q)=1$ and $F_{\rm DM}(q)=(\alpha\,m_e/q)^2$, respectively. For comparison, we show the expected sensitivity of a germanium target~\cite{Lee:2015qva} (with silicon performing similarly~\cite{Essig:2015cda}). As is evident, graphene can be competitive with the reach of semiconductor targets over the $\sim$MeV--GeV DM mass range, depending on the threshold energy.

\begin{figure*}[t]
\begin{center}
	\includegraphics[width=0.95\textwidth]{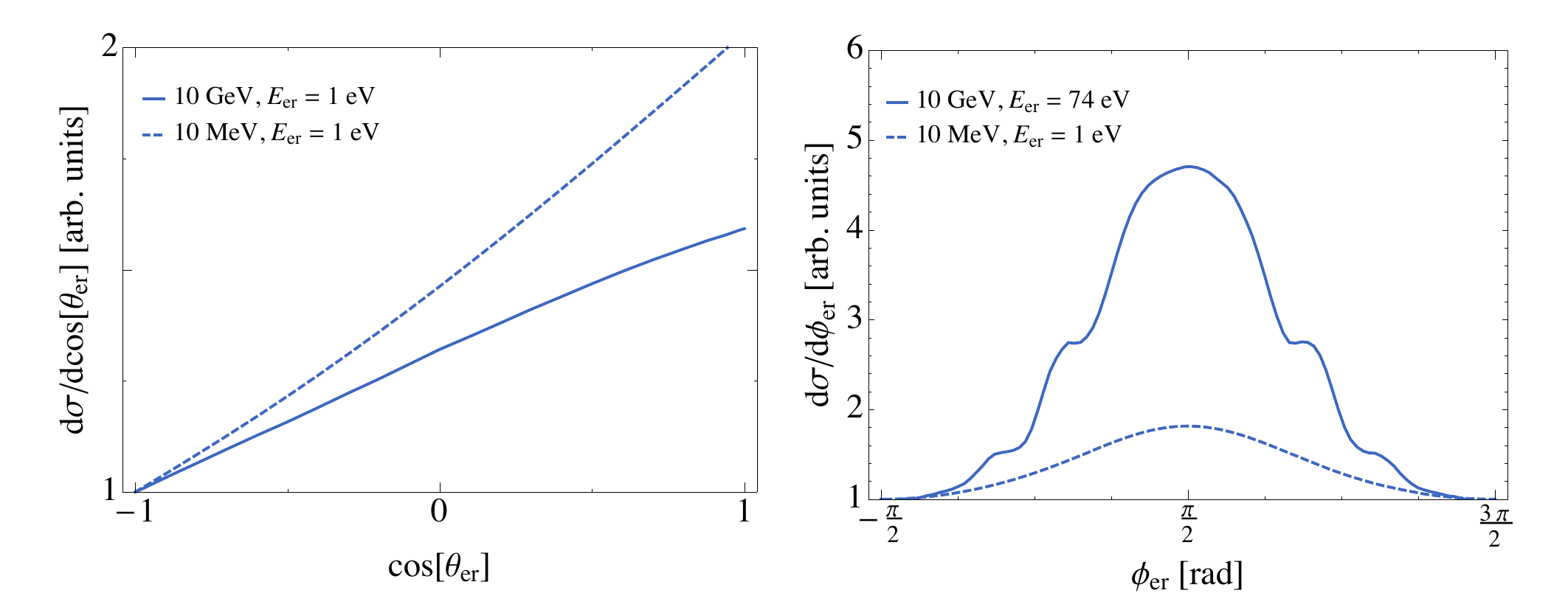}
	\vspace{-0.4cm}
\caption{Example angular distributions for DM masses 10~MeV (dashed) and 10~GeV (solid) in a DM stream with $v_{\rm stream} = 550$~km/s in the lab frame. ({\em left}) Polar distribution of the final-state electron when the stream is oriented perpendicular to the graphene plane and points in the $\hat z$ direction of $\cos\theta =1$, for $E_\text{er} = 1$~eV. ({\em right}) Azimuthal distribution of the final-state electron when the stream is oriented parallel to the graphene plane and points in the $\hat y$ direction of $\phi = \pi/2$. The results are shown for electron recoil of $E_\text{er} = 1$ (74)~eV for the 10~MeV (10~GeV) masses.
}
\label{fig:diffrac}
\end{center}
\vspace{-0.8cm}
\end{figure*}

\section{Directional Detection}\label{sec:dir}
In a 2D material, DM can scatter electrons directly into the vacuum without additional interactions. The electrons retain information about the initial direction of the DM, making 2D targets especially suitable for directional detection. In particular, the structure of the atomic orbitals implies that \emph{the outgoing electron is preferentially emitted in the direction of the incident DM.}  In Eq.~(\ref{eq:vdsigma}), the momentum-space orbital is evaluated at $\veck = \vecq - \veck_f$, and the appearance of $|\veck|^2$ in the denominator of Eq.~(\ref{eq:Phi2pz}) means the rate is maximized when $\vecq$ is as small as is kinematically allowed and $\veck_f$ is parallel to $\vecq$.\footnote{Note that $|\veck_f| \ll |\vecq|$ for typical kinematics \cite{Graham:2012su}, so the maximum and minimum values of $|\vecq - \veck_f|$ are typically the same order of magnitude, leading to $\mathcal{O}(1)$ differences in the forward versus backward rates.}  Solving the $\delta$-function in  Eq.~(\ref{eq:vdsigma}) then enforces that $\veck_f$ is parallel to $\mathbf{v}$ for these kinematics.  Identical arguments hold for the wavefunctions in the other bands.

To illustrate this behavior, we consider the angular distribution of the scattered electron in graphene for the case of a dispersion-less DM stream with $g(v) \propto \delta\left(v-v_\text{stream}\right)$ with $v_\text{stream} = 550$~km/s, for streams normal and parallel to the graphene plane.  The intuition afforded by these examples applies to generalized velocity distributions, which can always be broken down into parallel and normal components. A large stream velocity was chosen to make the azimuthal diffraction pattern more apparent, but the presence of the forward scattering peak is completely independent of the magnitude of the DM velocity.

The left panel of Fig.~\ref{fig:diffrac} shows the polar angular dependence of the scattering cross section for a DM stream normal to the plane, in the $\hat z$ direction. The curves are plotted for $E_\text{er} = 1$~eV for two DM masses, 10~MeV and 10~GeV. As anticipated, the differential rate is largest for forward scattering. Forward scattering is less favored for heavier DM because the minimum kinematically-allowed $q$ is smaller and the numerator of the $\pi$ wavefunction $\tilde{\phi}_{2\pz}(\vecq - \veck_f)$ is suppressed when $q \sim k_f$ (see \emph{e.g.}, Eq.~\eqref{eq:Phi2pz}). This may allow some rudimentary form of mass discrimination based on the ratio of forward to backward scattering rates.

The right panel of Fig.~\ref{fig:diffrac} shows the azimuthal dependence of the scattering cross section for a DM stream oriented parallel to the graphene plane, pointing in the $\hat y$ direction ($\phi=\pi/2$). Again, the electrons are preferentially emitted in the same direction as the stream, for both 10~GeV and 10~MeV DM masses.  The heavier DM curve is plotted for $k_f = 2\pi/a \simeq 8.7 \ {\rm keV}$ ($E_\text{er} \simeq 74 \eV$). A diffraction pattern is discernible in the angular distribution, arising from the interference between wavefunctions of neighboring carbon atoms.  The diffraction pattern is washed out if the velocity dispersion of the stream is greater than $\sim$25~km/s, but the scattering remains peaked in the forward direction.  For the lighter DM, $k_f\sim 2\pi/a$ cannot be achieved at an appreciable rate, so the differential distribution is shown for $k_f = 1 \ {\rm keV}$ ($E_\text{er} \simeq 1 \eV$).  While no diffraction pattern emerges for recoil momenta small compared to the inverse lattice spacing, a broad forward-scattering peak persists. For streams in different in-plane directions, the shapes of the forward-scattering peak and the secondary peaks change, but the general features remain the same.

We emphasize that, for the directional information of the ejected electron to persist, the electrons must exit the monolayer without significant rescattering.  Thus, a DM stream in the plane of the material should eject electrons at a sufficiently large angle from the plane, restricting the phase space for directional detection. For 50~eV electrons emitted at greater than $10^\circ$ from the plane, we estimate the angular spread due to rescattering to be $\sigma_\phi \sim 10^\circ$, with $\sigma_\phi$ decreasing for higher energy electrons. This is comparable to the width of the secondary diffraction peaks, but well below the width of the central forward-scattering peak, so we expect some diffraction structure to remain visible if there is a significant DM stream component.

\begin{figure*}[t] 
\begin{center}
	\includegraphics[width=0.8\textwidth]{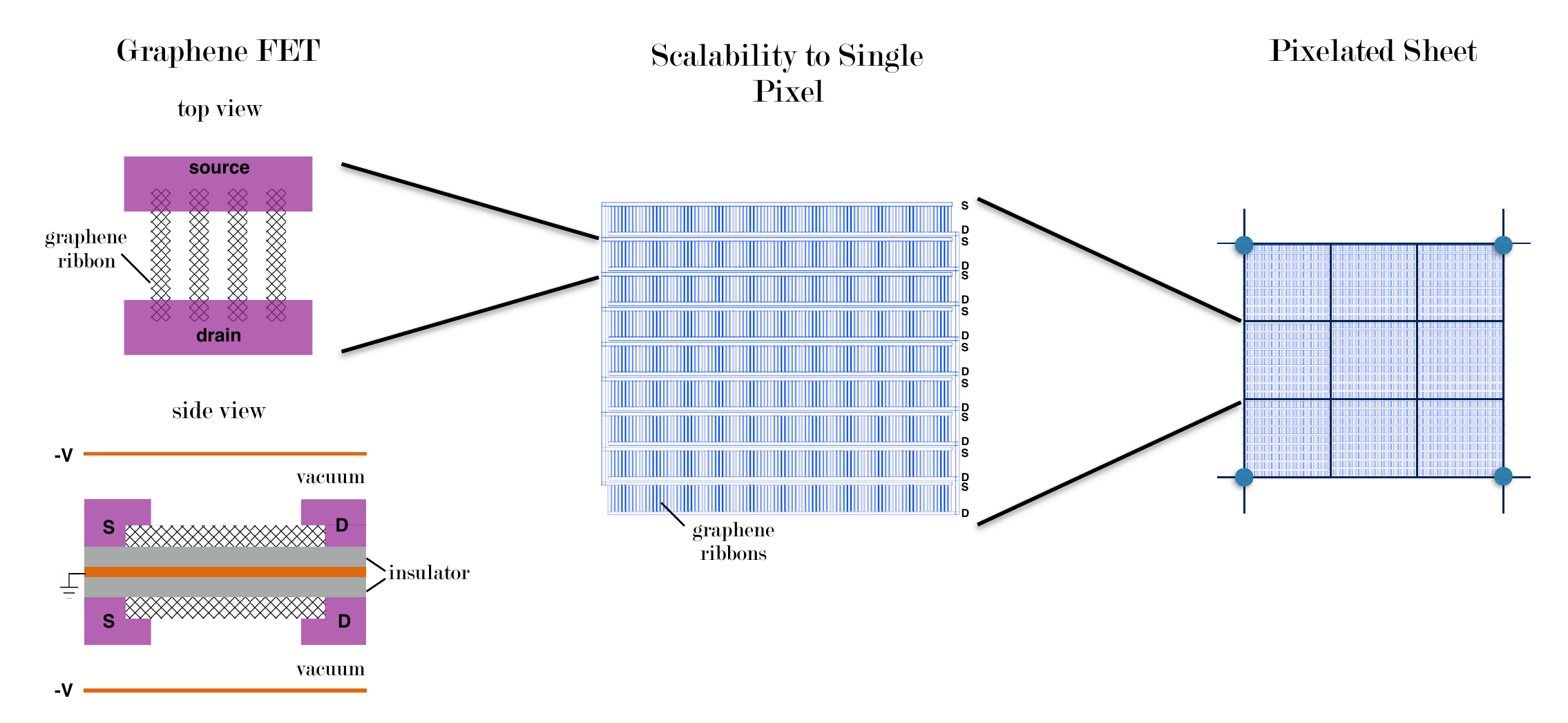}
	\vspace{-0.2cm}
   \caption{A graphene FET consists of a graphene ribbon grown on a substrate and connected to a source and drain (left).  The FET plane will be double-sided, separated by two insulating layers and a bottom gate electrode.  Top gate electrodes will provide the $\sim -100$~V needed to accelerate ejected electrons away from the electrodes and back towards the graphene planes.  Multiple graphene FETs can be arranged into a single pixel (center) with interdigitated source and drain.  The proposed experiment consists of stacked arrays of graphene sheets, where each sheet (right) consists of many individual pixels and is supported at the corners as shown in the diagram.}
   \label{fig:conceptdesign}
\end{center}
\vspace{-0.8cm}
\end{figure*}

Two-dimensional targets naturally allow for forward-backward discrimination of the DM direction, leading to a daily modulation in the event rate.  In particular, the experiment can be oriented such that the DM wind is nearly normal to the graphene planes twice a day, once parallel and once antiparallel.  In the proposed experimental configuration described in Section~\ref{sec:design} below, forward and backward electrons can be distinguished using a double-sided graphene pixel.  Simulating the full 3-dimensional DM velocity distribution $g(\mathbf{v})$ in the SHM, we find the rate in the forward direction is approximately a factor of 2 larger than in the backward direction for DM masses from 10~MeV to 10~GeV.  A daily modulation can be established at 95\% confidence with only $\sim$70 signal events, assuming zero background \cite{Lee:2013xxa}.

\section{Conceptual Experimental Design}
\label{sec:design}

We now outline a conceptual experimental design, along with a discussion of single-electron backgrounds. The experiment consists of pixelated graphene sheets, each grown onto a substrate, that are monitored for the ejection of an electron by virtue of the graphene/substrate system acting as a field-effect transistor (FET) \cite{Schedin:2007,schwierz2010graphene,sorgenfrei2011label,Sharf:2014}.  To obtain sufficient target mass in a compact volume, the graphene sheets are stacked and separated by vacuum.  When an electron is ejected from a pixel, an electric field drifts it either towards another FET within the detector volume, or towards a calorimeter at the boundary of the detector.  The combination of the position reconstruction of the electron and time-of-flight is sufficient to reconstruct the (fully directional) velocity of the ejected electron, with the energy measurement in the calorimeter providing an additional check on the kinematics.

The PTOLEMY experiment~\cite{Betts:2013uya} can realize this proposal with up to 0.5~kg of monolayer graphene, yielding competitive sensitivity to semiconductor targets. The primary goal of PTOLEMY is to detect electrons emitted from a tritium-loaded graphene surface after the capture of cosmic relic neutrinos.  If instead of holding tritium, the experiment is run using bare graphene surfaces, it is also sensitive to electrons ejected by DM scattering.

\subsection{Detector configuration}

The primary benefit of using a 2D target is that the scattered electron is ejected from the material into vacuum, at which point its trajectory can be manipulated by electric fields. The particular choice of graphene as the target is advantageous because the addition or removal of single electrons can cause measurable changes in the conductivity of graphene~\cite{Schedin:2007,schwierz2010graphene,sorgenfrei2011label,Sharf:2014}.  For example, the adhesion or desorption of molecules from graphene at room temperature causes single-electron changes in the local carrier density that manifests as a measurable change in resistivity~\cite{Schedin:2007}. At cryogenic temperatures, the resistivity change increases by an order of magnitude compared to room temperature, with even greater resistivity change possible by engineering the graphene-substrate system to open up a meV band gap \cite{novoselov2004electric}. As another example, carbon nanotube FETs can detect changes due to single electrons in their vicinity, again through changes in their conductivity~\cite{Sharf:2014}. Therefore, we imagine that each graphene ``pixel'' is coupled to a substrate in a FET configuration, so that the gate of the FET gets toggled whenever an electron is ejected, allowing one to identify that the pixel produced a hit. The same pixel FET may be used to detect incoming as well as outgoing electrons, allowing a coincidence measurement. We will require a coincidence measurement in exactly two FETs (or one FET and the outer calorimeter) for signal candidates, which will mitigate backgrounds from single FETs as well as from high-energy events which will trigger many FETs at once.

An example of the graphene FET is shown in the left panel of Fig.~\ref{fig:conceptdesign}.  Multiple FETs can be combined into a single pixel (Fig.~\ref{fig:conceptdesign}, center) with source and drain interdigitated to maximize the area covered by graphene.  A finite $\sim$meV band gap in the graphene will greatly increase the on/off current ratio of the FET.  This can be achieved via interactions with the substrate or with a ribbon structure.  The FET is back-gated to the neutrality point at the center of the band gap to minimize leakage across the source-drain.  For high purity graphene, the number of charge carriers in the channel is highly suppressed at low temperatures.  A single electron charge on the finite electrical capacitance of the ribbon produces a voltage step that increases the conductivity of the ribbon by many orders of magnitude, causing a macroscopic amount of charge to flow between source and drain.  The conduction across the source-drain, configured as an interdigitated capacitor, is read out at regular intervals and then reset. The only dead time in this setup is during FET readout, though each individual pixel is single-fire until readout. Note that the dark count from source-drain leakage in such a setup is negligible, as a slow voltage decay should easily be distinguishable from the sharp drop resulting from a change in conductivity. The work function of graphene also helps to suppress dark count from ejected electrons.

Achieving a sufficient target volume of $\sim$0.5~kg of \linebreak graphene requires $10^{10} \ {\rm cm}^2$ of surface area. To fit inside a compact volume, the surface area should be divided into pixelated sheets (Fig.~\ref{fig:conceptdesign}, right) stacked in an array and supported at the corners with epoxy or a similar material.\footnote{A similar modular design has been used in PANDA-X III \cite{Chen:2016qcd}.} For pixel areas of $1 \ {\rm mm}^2$, $10^4$ pixels per sheet (wafer), and vertical separations of $\sim$mm in $\hat z$, the entire target volume can fit inside a $\sim$10$^3$~m$^3$ space. In a preliminary study, we have explored electric field configurations that allow for efficient electron transport in this compact volume. We show a conceptual design in Fig.~\ref{fig:detect}~(left). In the innermost detector volume, we impose an electric field which is mostly normal to the sheets to repel electrons from the conducting planes and direct them back into neighboring pixels in the same plane.
A potential difference of 100~V, corresponding to a maximum $E$-field of 100~V/mm, is sufficient to repel electrons of energy less than 100 eV. In the detector volume closer to the boundary, we impose an electric field which is mostly parallel to the sheets to direct electrons to a calorimeter at the boundary. The calorimeter offers the additional advantage of being able to measure the electron energy; a full design would optimize between the two detection modalities to ensure that the maximum kinematic information is kept for each event. The outermost volume is fiducialized in order to reduce backgrounds. This setup is reminiscent of a time-projection chamber (TPC), which is the technology of choice for current directional detectors~\cite{Mayet:2016zxu}.  In a TPC, secondaries of the scattering interaction are drifted through a gas target towards a segmented anode, where they are detected.  Our proposal is similar, however the primary electron is now deflected to the detection site through vacuum.

\subsection{Directionality}

\begin{figure*}[t] 
\begin{center}
   \includegraphics[width=0.85\textwidth]{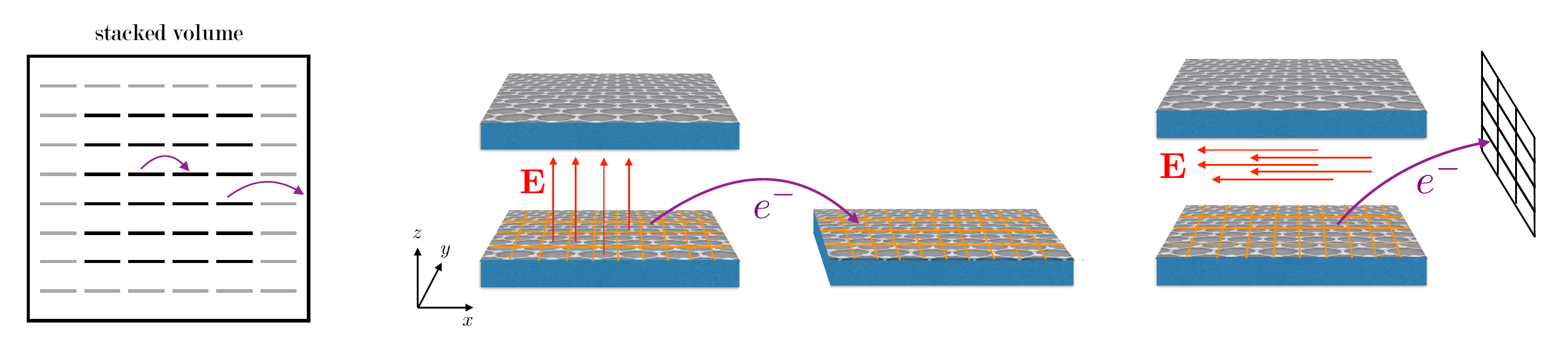}
   \caption{A conceptual design for graphene directional detection.  The left panel illustrates a cut-out of the stacked volume of graphene sheets that form the detector.  For graphene sheets in the inner detector volume, scattered electrons follow a ``FET-to-FET'' trajectory (center panel).  In this case, an electron in the center of the detector is repelled from the layer above by a perpendicular $E$-field and drifts ballistically to a neighboring pixel (orange squares).  For graphene sheets near the detector edges, electrons follow a ``FET-to-calorimeter'' trajectory (right panel).  In this case, an electron near the outside of the detector is drifted by a parallel $E$-field to a segmented calorimeter.  A full design of the experiment would combine the FET-to-FET and FET-to-calorimeter modalities in an optimal way.  To reduce background contamination, one can only consider scattering events from within a fiducialized volume (denoted by the black lines, left panel), ignoring events that originate on the outermost sheets in the detector volume (gray lines).}
   \label{fig:detect}
\end{center}
\vspace{-0.8cm}
\end{figure*}

In the setup we have described, a daily modulation of the count rate is essentially automatic, with no spectral or velocity information required. The reason for this is that in a double-sided pixel FET, electrons scattered upward from the top FET will escape to vacuum, while electrons scattered upward from the bottom FET will hit the substrate and are vetoed due to lack of a coincident hit in either another FET or the calorimeter. The reverse is true for electrons scattered downwards. Over the course of a day, the direction of the DM wind changes relative to the graphene sheet.  If the DM wind is oriented along the $\hat z$-direction of the experiment (the upward-pointing normal to the graphene planes), a coincidence signal will primarily be seen from the top FET layers where electrons can escape to vacuum. Twelve hours later when the orientation is reversed, a coincidence signal will primarily be seen in the bottom FET layer. The top and bottom layer are separated by a grounded electrode and there should be no cross-talk between the two layers.  Thus, forward-backward discrimination of the electron (or `head-tail' discrimination~\cite{Mayet:2016zxu}), a key feature of a DM signal, is inherent in this design.  As described in Section~\ref{sec:dir}, we have simulated this effect incorporating the full velocity distribution of the SHM, and found that the forward rate is approximately twice as large as the backward rate for DM masses from 10~MeV to 10~GeV. The ratio of top-layer FET rates to bottom-layer rates should track this modulation.


A more refined determination of the DM's direction can be achieved if the pixel FET response is fast enough compared to the electron drift time to obtain time-of-flight information. The electrons are nonrelativistic, with speeds on the order of $10^{6}$--$10^{7}$~m/s, corresponding to drift times of $\sim$ns, which are feasible to achieve with graphene FETs. Information on the directionality of the signal provides a powerful discriminant against backgrounds, and also opens the possibility of mapping the local DM phase space directly. For electrons in the inner volume, full 3-dimensional reconstruction of the initial velocity can be obtained by knowing the coordinates of the starting and ending pixels (Fig.~\ref{fig:detect}, center), as well as the time of flight $\Delta t$. As a simple example, in a configuration with purely normal $E$-field, the vertical velocity $v_z$ can be determined from solving $\Delta z = -\frac{1}{2m_e}eE (\Delta t)^2 + v_z \Delta t$, where we require $\Delta z = 0$ for the electron to start and end on the same layer, as shown in the sample trajectory in Fig.~\ref{fig:detect} (left). Since the electron drifts ballistically in $x$ and $y$, time-of-flight gives $v_x = \Delta x/\Delta t$ and $v_y = \Delta y/\Delta t$. 
The electron energy can be recovered from the initial velocity vector, depending on the relative uncertainties of the three velocity components; a full analysis of the velocity and energy resolution requires a dedicated simulation.

For electrons in the outer volume (Fig.~\ref{fig:detect}, right), 3-dimensional velocity reconstruction is also possible, with the energy measurement from the calorimeter replacing the time-of-flight measurement. We expect to be able to achieve $\sim 1$~eV energy resolution by scaling up existing measurements for single-IR photon counting \cite{lita2008counting}, which has demonstrated resolutions of 0.29~eV for a 0.8~eV single photon.  Alternatively, lower calorimeter resolution with a pixelated FET array instrumented on top of a low dark-current cryogenic CCD array may be acceptable if combined with the higher resolution information from the FET-to-FET time-of-flight.

In order to prevent rescattering of the primary electron, the whole experiment must be in a high-vacuum environment. A pressure of $2 \times 10^{-7}$~torr corresponds to a mean free path for electrons of roughly 500 m, which is more than sufficient for a $\sim$10~m$\times$10~m$\times$10~m target volume where electron trajectories are expected to be $\sim$cm in length. We expect this vacuum level to be technically feasible during the assembly of the target volume, as KATRIN has already achieved $10^{-11}$~torr in a 1042~m$^3$ volume~\cite{wolf2009size}.  Each FET plane will be vacuum sealed on top and bottom during assembly, similar to the method described in Ref.~\cite{huebner2015high}.  The large vacuum volume is relevant for the regions outside of the sealed planes at the boundaries of the target volume.  The target will be kept at cryogenic temperatures and have no line-of-sight vacuum trajectories from the outer vacuum region to the sealed FET planes.  Residual gas backgrounds will be cryopumped to the outer boundaries of the fiducialized volume.  We expect the quality of the vacuum inside the target volume to be sufficient to operate the experiment for prolonged exposure periods without having to reopen the target.

\subsection{Overburden}
With an area per plane of $10^{6} \ {\rm cm}^2$, the overburden of cosmic-ray muon flux is an important concern for
dead-time associated with a cosmic-ray veto.  The instrumented target is designed to have no more than a
percent-level fill factor of support material, mostly epoxy or a similar material to support the graphene sheets at the corners as shown in Fig.~\ref{fig:conceptdesign} (right).  The remainder of the target volume will be highly sensitive to charged
particles entering the volume, and therefore the electric field regions that control the conductivity of the 
graphene FETs, including the regions between the vacuum-separated top gate electrode and the graphene and 
underneath the graphene with the insulator-separated bottom gate electrode, will be active regions for cosmic-ray vetos.  With an overburden of roughly 3~km or greater, as would be the case for an underground lab like Gran Sasso or SNOLAB, the total flux of muons across the entire graphene target falls below $10^{-1} \ {\rm s}^{-1}$ \cite{Mei:2005gm}.  With a finite readout 
time of the FET planes, this rate would introduce less than 1\% of dead-time depending to a lesser extent on the size 
of the fiducialized volume used in the veto.

\subsection{Single-electron backgrounds}
Any incident particle with sufficient energy to eject a valence electron can in principle pose a background. One of the primary backgrounds for such an experiment comes from environmental radioactivity, which can be mitigated by shielding, cryopumping, and the use of materials of high radiopurity with specialized fabrication techniques~\cite{Chen:2016qcd,jastram2015cryogenic}. In particular, the substrate can be a source of radioactive backgrounds, but this can be minimized by using an atomically thin substrate with atomic-layer depositions of aluminum oxide, or a thicker substrate with fewer radioisotopes. Many of the standard techniques used to control such backgrounds in current direct detection experiments will be efficient here as well, because the experimental design acts as an \emph{effective} three-dimensional volume.  For example, highly energetic products from radioactive decays may traverse multiple graphene layers, knocking out several electrons.  This would be recorded as a multiple-scatter event, which can be vetoed. A background event ejecting only a single secondary electron in the outermost layers of the target volume would be more difficult to distinguish from signal.  However, fiducialization of the volume can aid in reducing such contributions.

We expect the main irreducible background to be $^{14}$C decay in the graphene.  The landmark work on isotope enrichment used in 1991 for the Borexino experiment to assess the content of $^{14}$C in methane in natural gas achieved levels of $\lesssim 1.6 \times 10^{-18}$, relative to $^{12}$C~\cite{Mosteiro:2015boa}.  Accelerator mass spectroscopy during the fabrication process may be able to reduce the $^{14}$C fraction from the $10^{-18}$ achieved by Borexino to $10^{-21}$ \cite{litherland2005low}.  Graphene grown with cold plasma techniques developed at the Princeton Plasma Physics Laboratory with $\lesssim10^{-21}$ levels of $^{14}$C  will leave $\sim$$10^4$ atoms of $^{14}$C in the 0.5 kg of graphene, with a half-life of 5700 years. This translates into roughly 1-2 events per year assuming no veto, which is already at a negligible level. However, further reduction is expected with veto power, which may be necessary should the required radiopurity be difficult to achieve. 

To understand the expected veto efficiency, we have completed a GEANT4~\cite{Agostinelli:2002hh,Allison:2006ve} simulation of the interaction of the $^{14}$C beta with a 0.34~nm thick graphene sheet. We find that outgoing betas emitted at angles above $\sim 10^\circ$ from the plane of the graphene deposit less than 2--10~eV independent of the beta energy, which could result in a secondary electron with kinetic energy below 5 eV being emitted. We estimate the veto efficiency as a function of the $^{14}$C beta energy to be highly efficient above 1~keV, and therefore, the background is suppressed by $\sim$$10^{-2}$ from the veto, which further reduces the contribution from $^{14}$C. This could also be combined with a cut on the electron energy to veto these low-energy secondaries. However, betas emitted nearly coplanar with the sheets may deposit significant amounts of energy, and will likely pose an irreducible background, highlighting the need to achieve radiopurity of $10^{-21}$.

The general arguments presented here are meant to demonstrate the feasibility of our experimental proposal.  A more detailed design is currently under development, along with a careful consideration of single-electron backgrounds.


\section{Conclusions}\label{sec:concl}
This Letter presents a proposal for directional DM detection with 2D targets.  If sufficient energy is deposited by the DM scattering, the electron can be ejected from the target and detected on another graphene sheet or in a calorimeter.  The electron retains information about its recoil direction, which is in turn correlated with the incoming DM direction.  For a graphene target, this setup, which can be implemented by PTOLEMY, can probe DM down to MeV masses. The reach is comparable to that for semiconductor targets, with the added benefit of directionality.  Further improvement can be made by lowering the graphene work function. Other 2D materials, such as monolayer gold \cite{Drukier:2012hj}, could be similarly powerful.

Advantageously, the same experiment can also be used to detect nuclear recoils, similar to proposals that have been made for the ejection of carbon ions from nanotubes~\cite{Capparelli:2014lua,Cavoto:2016lqo}.   Only $\sim$20~eV of energy is needed to eject a carbon atom from the graphene sheet (and slightly more to eject an ion)~\cite{displacement}.  The ion can either be detected calorimetrically or by monitoring the conductivity of the graphene (see \emph{e.g.},\ \cite{schwierz2010graphene,sorgenfrei2011label}).  This would enable the same experimental setup to probe nuclear scattering down to $\sim$GeV DM masses.

Lastly, 2D materials with a small band gap, such as graphene, may be sensitive to DM as light as the warm DM limit of $\sim$keV masses.
While small energy gaps and large target electron velocities have already been shown to allow superconductors to probe keV DM~\cite{Hochberg:2015pha,Hochberg:2015fth}, this target also exhibits a large optical response.  Superconductors are thus limited in their sensitivity to scattering processes mediated by dark photons~\cite{Hochberg:2015fth}.
As a result, a material like graphene, with a weaker optical response, can be highly complementary to a superconducting target.  We leave a detailed study of 2D targets for detection of keV--MeV DM for future work.

\vspace{0.25in}
\mysection{Acknowledgments} We thank Timothy Berkelbach, Garnet Chan, Youngkuk Kim, Andrew Rappe, and Joe Subotnik for enlightening discussions regarding graphene. We also thank Snir Gazit, Adolfo Grushin, Roni Ilan, Aaron Manalaysay, Dan McKinsey, Antonio Polosa, Matt Pyle, and Zohar Ringel for conversations about light dark matter detection in various materials. YH is supported by the U.S. National Science Foundation under Grant No.
PHY-1002399 and Grant No. PHY-1419008, the LHC Theory Initiative. ML is supported by the DoE under contract DESC0007968, as well as by the Alfred P. Sloan Foundation. CT is supported by the Simons Foundation (\#377485) and John Templeton Foundation(\#58851). KZ is supported by the DoE under contract DE-AC02-05CH11231.

\newpage
\begin{appendix}
\section{Electron Wavefunctions}
\label{app:Wavefunctions}

This Appendix reviews the calculation of the analytic forms for the $\pi$ and $\sigma$ electron wavefunctions in graphene in the tight-binding approximation, which are needed for the determination of the scattering rate in Eq.~\eqref{eq:vdsigma} of the main text.  Our discussion follows Chap.~2 of Ref.~\cite{dresselhaus} closely and we refer the interested reader there for a more comprehensive introduction.
\begin{figure*}[t] 
     	\begin{center}
	 \includegraphics[width=0.8\textwidth]{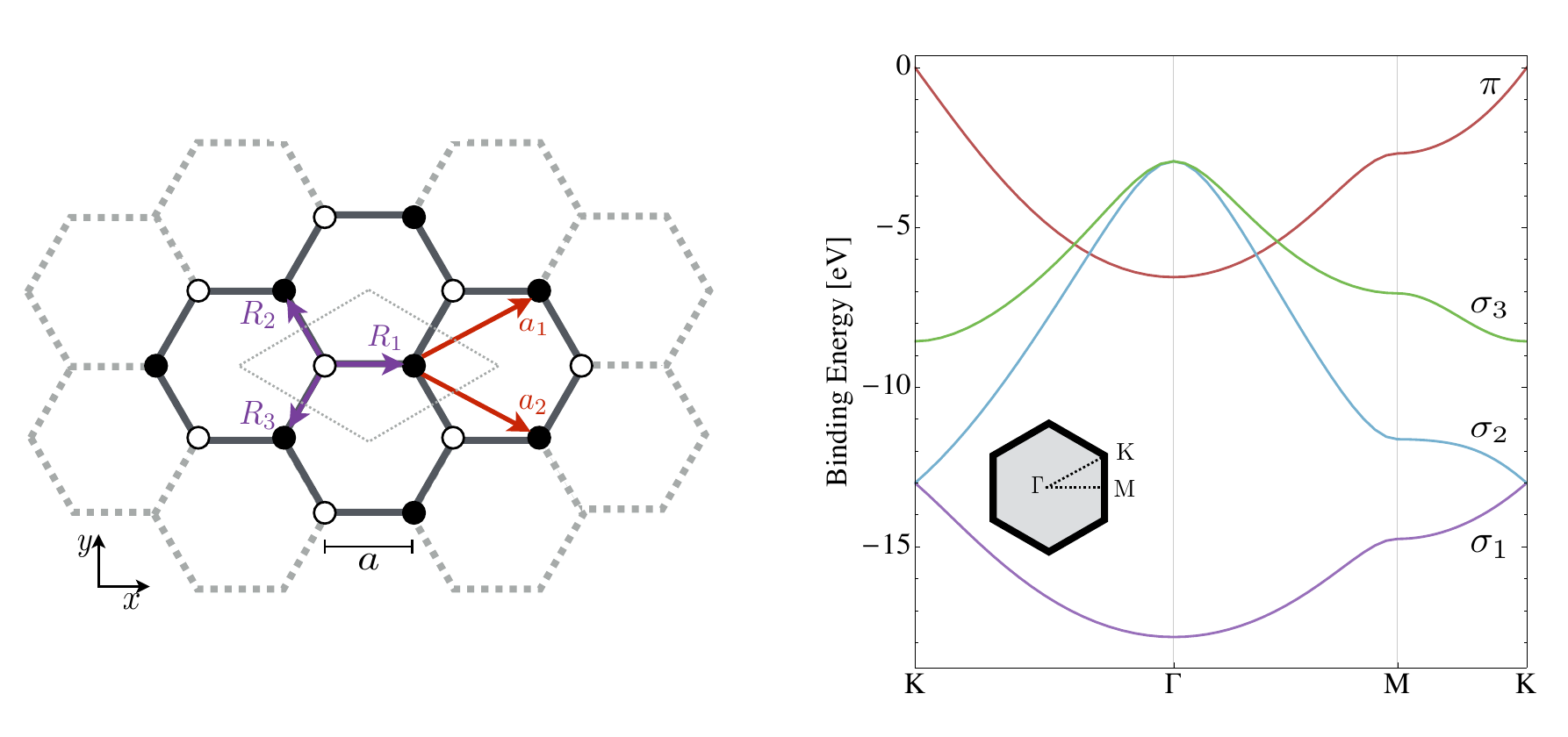}
	\end{center}
	   \vspace{-0.25in}
   \caption{(\emph{left}) Graphene is comprised of two triangular carbon sub-lattices, which are illustrated by the open and solid circles.  The lattice vectors $\mathbf{a}_{1,2}$ are indicated by the red arrows, and the nearest-neighbor vectors $\mathbf{R}_{1,2,3}$ are shown in purple.  The gray diamond depicts the unit cell.  The nearest-neighbor distance is $a=0.142$~nm.  (\emph{right})  The valence-band diagram for graphene, as determined from the procedure outlined in the Supplementary Material.  The Brillouin zone is shown in the inset, with the high-symmetry points $\Gamma$, K, and M labeled.  }
  \label{fig:graphene}
\end{figure*}

The tight-binding Bloch wavefunction for an electron located on sub-lattice $\gamma$ with periodic lattice momentum $\vecl=(\ell_x, \ell_y)$ is
\be
\Phi_{\gamma} (\vecl, \vecr) = \frac{1}{\sqrt{N}} \sum_{N} e^{i \vecl \cdot \mathbf{R}_N} \, \phi_s\left(\vecr- \mathbf{R}_N\right),
\label{eq:Bloch}
\ee
where the sum runs over $N$ lattice sites $s$ with position $\mathbf{R}_N$.  In practice, we only sum over nearest-neighbor sites.  For example, if the coordinate system is centered at an A site in the unit cell (open circle in left panel of Fig.~\ref{fig:graphene}), then the nearest neighbors consist of three B sites (solid circles) located at $\mathbf{R}_{1}, \mathbf{R}_2$, and $\mathbf{R}_3$.  The Bloch wavefunctions are then
\begin{eqnarray}
\Phi_A(\vecl, \vecr) &=& \phi_A\left(\mathbf{r} \right) \\ \nonumber
\Phi_B(\vecl, \vecr) &=& \sum_{j = 1}^{3} e^{i \vecl\cdot\mathbf{R}_j} \phi_B(\vecr - \mathbf{R}_j) \, .
\label{eq:BlochAB}
\end{eqnarray}
Electrons in the $\pi$ band are in the $2\pz$ atomic orbital of carbon.  Therefore, $\phi_A (\vecr) = \phi_B(\vecr) = \phi_{2\pz}(\vecr)$ because the A and B sites are both identical carbon atoms.  The hydrogenic orbital for a $2\pz$ electron in carbon is
\be\label{eq:orb}
\phi_{2\pz}(\vecr) = 3.23\, a_0^{-3/2}\, \frac{r}{a_0}\, e^{- Z_\text{eff} r/2 a_0}\, \cos \theta\,,
\ee
with $a_0$ the Bohr radius. We take the effective charge $Z_\text{eff} \simeq 4.03$ to reproduce the nearest-neighbor overlap $s=0.129$~\cite{dresselhaus}.
The wavefunction for a $\pi$ electron in graphene is therefore
\be
\Psi_\pi(\vecl, \vecr) = C_A(\vecl) \, \Phi_A(\vecl, \vecr) + C_B(\vecl) \, \Phi_B (\vecl, \vecr) \, ,
\ee
where $C_{A,B}$ are constants that depend on $\vecl$.  In the following, we will suppress their $\vecl$ dependence for notational simplicity.  This wavefunction must satisfy
\be
H \, \Psi_\pi = E_\pi(\vecl) \, \Psi_\pi \, ,
\label{eq:Hamiltonian}
\ee
where $H$ is the crystal Hamiltonian and $E_\pi$ is the energy of the $\pi$-band electron.  Defining the vector $\mathbf{C} = (C_A, C_B)$, Eq.~\eqref{eq:Hamiltonian} can be rewritten as
\be
\mathcal{H} \, \mathbf{C} = E_\pi(\vecl) \, {\cal S} \, \mathbf{C} \, .
\ee
Here, ${\cal S}$ is the overlap matrix \begin{equation}
{\cal S}_{\gamma \gamma'} = \left\langle \Phi_\gamma(\vecl, \vecr) | \Phi_{\gamma'}(\vecl, \vecr)\right\rangle \, ,
\end{equation}
which becomes
\be
{\cal S} =  \left(
\begin{array}{cc}
1 & s f(\vecl) \\
s f(\vecl)^* & 1
\end{array} \right) \,
\ee
for the $\pi$ electrons.  The overlap between nearest-neighbor atomic orbitals is
\begin{equation}
s = \int d^3 r \, \phi^*_{2\pz}(\vecr) \, \phi_{2\pz}(\vecr - \mathbf{R}_j) = 0.129 \, .
\label{eq:overlap}
\end{equation}
Additionally, $f(\vecl)$ is the phase factor
\begin{eqnarray}
f(\vecl) & = & e^{i \vecl \cdot \mathbf{R}_1} +e^{i \vecl \cdot \mathbf{R}_2} + e^{i \vecl \cdot \mathbf{R}_3} \\ \nonumber
 & = & e^{i \ell_x a} + 2 e^{-i \ell_x a/2} \cos \left(\frac{\sqrt{3}\ell_y a}{2}\right)\, ,
\end{eqnarray}
where $a=0.142$~nm is the nearest-neighbor distance.

Similarly, the transfer matrix $\mathcal{H}$ is defined in terms of the crystal Hamiltonian $H$ as
\begin{equation}
{\cal H}_{\gamma \gamma'} = \left\langle \Phi_\gamma(\vecl, \vecr)\right| H \left|\Phi_{\gamma'}(\vecl, \vecr) \right\rangle \, ,
\end{equation}
and can be written as
\be
{\cal H} =  \left(
\begin{array}{cc}
 \epsilon_{2\p}& t f(\vecl) \\
t f(\vecl)^* &  \epsilon_{2\p}
\end{array} \right) \, ,
\ee
where the transfer integral $t$ is
\be
t = \int d^3 r \, \phi^*_{2\pz}(\vecr) \, H \, \phi_{2\pz}(\vecr - \mathbf{R}_j) = -3.03~\text{eV} \, .
\ee
By convention, the $2\p$ orbital energy is set to the reference value $\epsilon_{2\p}= 0$.

The energy eigenvalues are obtained by solving $\det\left[ \mathcal{H} - E_\pi(\vecl) \cal{S} \right] = 0$.
The result is
\be
E_\pi (\vecl) = \frac{\epsilon_{2\p} \pm t |f(\vecl)|}{ 1 \pm s  |f(\vecl)|} \, ,
\ee
where the plus sign corresponds to the valence $\pi$ band and the minus sign corresponds to the conduction $\pi^*$ band.  The corresponding eigenvectors are
\be
\mathbf{C}(\vecl) = \frac{1}{\sqrt{2}} \left( \begin{array}{c}
1 \\
\pm e^{i \varphi_\vecl}  \end{array} \right) \, , \text{ with } \varphi_\vecl = -\arctan \left( \frac{\text{Im} f(\vecl)}{\text{Re} f(\vecl)}\right) \, .
\ee
Thus, the complete expression for the $\pi$ wavefunction is
\be
\Psi_\pi(\vecl, \vecr) = \mathcal{N}_\vecl \left( \Phi_A (\vecl, \vecr)  + e^{i \varphi_\vecl} \, \Phi_B(\vecl, \vecr) \right) \, ,
\ee
where $\mathcal{N}_\vecl$ is a normalization constant.
For our rate calculation, we must also consider the $\sigma$ electrons, which are in a superposition of the
$2\s$, $2\px$, and $2\py$ orbitals. Since these orbitals are even under reflection in the graphene plane while the $2\pz$ orbital is odd, there is a super-selection rule forbidding the $\pi$ band from mixing with the $\sigma$ bands, and we can diagonalize the $\sigma$ electrons separately. To obtain the energies and wavefunctions for the $\sigma$ electrons, we follow a similar procedure as for the $\pi$ electrons.  However, the calculation is now more involved as the transfer and overlap matrices are $6\times6$ when written in the $\left(2\s^A,\,2\px^A,\,2\py^A,\,2\s^B,\,2 \px^B,\,2\py^B\right)$ basis.  For example, the $AA$ sub-matrices are
\be
{\cal S}_{AA} =  \left(
\begin{array}{ccc}
1 & 0 & 0\\
0 & 1 & 0 \\
0 & 0 & 1
\end{array} \right)\text{ and } {\cal H}_{AA} =  \left(
\begin{array}{ccc}
\epsilon_{2\s} & 0 & 0 \\
0 &\epsilon_{2\p}& 0 \\
0 &  0 & \epsilon_{2\p} \\
\end{array} \right) \, ,
\label{eq:AAsub}
\ee
where $\epsilon_{2\s} = -8.87$~eV is the energy of the $2s$ orbital relative to $\epsilon_{2\p} = 0$.  The $BB$ sub-matrices are identical to Eq.~\eqref{eq:AAsub}. The $AB$ sub-matrices are more complicated: for the overlap matrix, we have
\be
{\cal S}_{AB} =  \left(
\begin{array}{ccc}
{\cal S}_{\s\s} & {\cal S}_{\s\px} & {\cal S}_{\s\py} \\
-{\cal S}_{\s\px}& {\cal S}_{\px\px}  & {\cal S}_{\px\py} \\
-{\cal S}_{\s\py} & {\cal S}_{\px\py} & {\cal S}_{\py\py}
\end{array} \right) \, ,
\label{eq:ABsub}
\ee
with elements
\begin{eqnarray}
{\cal S}_{\s\s} & = & S_{\s\s}\left( e^{i \ell_x a} + 2 e^{-i \ell_x a/2} \cos \left(\frac{\sqrt{3}\ell_y a}{2}\right)\right) \nonumber \\
{\cal S}_{\s\px} & = & S_{\s\p}\left( -e^{i \ell_x a} +  e^{-i \ell_x a/2} \cos \left(\frac{\sqrt{3}\ell_y a}{2}\right)\right) \nonumber \\
{\cal S}_{\s\py} & = & -i \sqrt{3} \,S_{\s\p} \,e^{-i \ell_x a/2} \sin \left(\frac{\sqrt{3}\ell_y a}{2}\right) \nonumber \\
{\cal S}_{\px\px} & = & -S_\sigma \,e^{i \ell_x a}  + \frac{(3 S_\pi-S_\sigma)}{2} e^{-i \ell_x a/2} \cos \left(\frac{\sqrt{3}\ell_y a}{2}\right) \nonumber \\
{\cal S}_{\px\py} & = & \frac{i \sqrt{3}}{2} (S_\sigma + S_\pi) e^{-i \ell_x a/2} \sin \left(\frac{\sqrt{3}\ell_y a}{2}\right)\nonumber \\
{\cal S}_{\py\py} & = & S_\pi \, e^{i \ell_x a}  + \frac{(S_\pi-3S_\sigma)}{2}  e^{-i \ell_x a/2} \cos \left(\frac{\sqrt{3}\ell_y a}{2}\right) \, . \nonumber
\end{eqnarray}
The elements of the $AB$ transfer matrix, $\mathcal{H}_{AB}$, can be constructed by replacing ${\cal S} \rightarrow {\cal H}$ and $S \rightarrow H$ in Eq.~\eqref{eq:ABsub}.

\begin{table*}
\begin{tabular}{cccc} \hline
${\cal S}$~~ & ~~value~~ & ~~${\cal H}$~~ & ~~value (eV)~~  \\ \hline
$S_{\s\s}$ & 0.21 & $H_{\s\s}$ & -6.77 \\
$S_{\s\p}$ & 0.16 & $H_{\s\p}$ & -5.58 \\
$S_{\sigma}$ & 0.15 & $H_{\sigma}$ & -5.04 \\
$S_{\pi}$ & 0.13 & $H_{\pi}$ & -3.03  \\ \hline
\end{tabular}
\caption{Inputs for the transfer and overlap matrices~\cite{dresselhaus}.}
\label{tab:values}
\end{table*}

Table~\ref{tab:values} shows the numerical values of the transfer and overlap matrices used in our calculation. $S_{\s\s}, S_\sigma,$ and $S_\pi$ are taken from Ref.~\cite{dresselhaus}.  Because the graphene plane breaks spherical symmetry, the $Z_{\rm eff}$ used in the $2\pz$ orbital need not be the same as for the other orbitals. For self-consistency, we choose the values of $Z_{\rm eff}$ for the $2\px/2\py$ and $2\s$ orbitals separately to reproduce the values for $S_{\sigma}$ and $S_{\s\s}$ in Table~\ref{tab:values}: $Z_{\rm eff}^{2\px/2\py} \simeq 5.49$ and $Z_{\rm eff}^{2\s} \simeq 4.84$. This fixes the functional form of all valence orbitals:
\begin{align}
\phi_{2\px} (\vecr) & = {\cal N} \,a_0^{-3/2}\, \frac{r}{a_0}\, e^{- Z_{\rm eff}^{2\px/2\py} r/2 a_0}\, \sin \theta\, \cos \varphi, \\
\phi_{2\py} (\vecr) & = {\cal N} \,a_0^{-3/2}\, \frac{r}{a_0}\, e^{- Z_{\rm eff}^{2\px/2\py} r/2 a_0}\, \sin \theta\, \sin \varphi, \\
\phi_{2\s}(\vecr) & = {\cal N} \,a_0^{-3/2} \left(1 - \frac{Z_{\rm eff}^{2\s} r}{a_0} \right) e^{- Z_{\rm eff}^{2\s} r/2 a_0}.
\end{align}

The resulting self-consistent value for the $2\s/2\p$ overlap $S_{\s\p}$ is 0.163, which differs slightly from $S_{\s\p} = 0.10$ found in Ref.~\cite{dresselhaus}. However, the uppermost $\sigma$ band is unaffected by this change, and the effect on the lower two $\sigma$ bands is at most 1 eV, so we do not expect this difference to appreciably affect the rate. The band structure for our choice of parameters is plotted in the right panel of Fig.~\ref{fig:graphene}.

The momentum space orbitals are well-approximated by
\begin{align}
\widetilde{\phi}_{2\px}(\veck) & \approx \widetilde{{\cal N}} \, a_0^{3/2} \frac{a_0 \,k_x }{\left( a_0^2 \, |\veck|^2+ \left(Z_{\rm eff}^{2\px/2\py}/2\right)^2\right)^3}, \\
\widetilde{\phi}_{2\py}(\veck) & \approx \widetilde{{\cal N}} \, a_0^{3/2} \frac{a_0 \,k_y }{\left( a_0^2 \, |\veck|^2+ \left(Z_{\rm eff}^{2\px/2\py}/2\right)^2\right)^3}, \\
\widetilde{\phi}_{2\s}(\veck) & = \widetilde{{\cal N}} \, a_0^{3/2} \frac{a_0^2 |\veck|^2 - \left(Z_{\rm eff}^{2\s}/2\right)^2 }{\left( a_0^2 \, |\veck|^2+ \left(Z_{\rm eff}^{2\s}/2\right)^2\right)^3}.
\end{align}
The 2s result is exact as the Fourier transform can be performed analytically. In all cases, the dependence on $|\veck|^2$ in the denominator dominates over the smaller powers of $\veck$ in the numerator, and thus the argument for forward scattering described in the main text holds equally well for all four bands.

\end{appendix}

\bibliography{graphene}

\end{document}